\documentstyle{cupconf}
\input psfig.tex

\ifoldfss
\else
  \ifnfssone
    \newmathalphabet{\mathit}
      \addtoversion{normal}{\mathit}{cmr}{m}{it}
      \addtoversion{bold}{\mathit}{cmr}{bx}{it}
    \newmathalphabet{\mathcal}
      \addtoversion{normal}{\mathcal}{cmsy}{m}{n}
    \else
    \ifnfsstwo
    \fi
  \fi
\fi

\def \lta {\mathrel{\vcenter
     {\hbox{$<$}\nointerlineskip\hbox{$\sim$}}}}
\def \gta {\mathrel{\vcenter
     {\hbox{$>$}\nointerlineskip\hbox{$\sim$}}}}
\def \m{\ifmmode M_\odot\else M$_\odot$\fi}
\def \r{\ifmmode R_\odot\else R$_\odot$\fi}
\def\cms{cm$^{-1}$}
\def\kms{km$^{-1}$}
\def\gmcm3{gm~cm$^{-3}$}
\def\gm-s{gm~s$^{-1}$}
\def\cm3s{cm$^3$~s$^{-1}$}
\def\kms{km~s$^{-1}$}
\def\erg-s{erg~s$^{-1}$}

\def\beq{\begin{equation}}
\def\eeq{\end{equation}}
\def\ref{\reference}
\def\gr{$\gamma$-ray}
\def\grs{$\gamma$-rays}
\def\grb{$\gamma$-ray burst}
\def\grbs{$\gamma$-ray bursts}
\def\0{\parindent=0cm}
\def\5{\parindent=.5cm}
\def\7{\parindent=.7cm}
\def \lta {\mathrel{\vcenter
     {\hbox{$<$}\nointerlineskip\hbox{$\sim$}}}}
\def \gta {\mathrel{\vcenter
     {\hbox{$>$}\nointerlineskip\hbox{$\sim$}}}}
 \def\ni{$^{56}{\rm Ni}\ $}      
 \def\co{$^{56}{\rm Co}\ $}      
\def\hexnumber#1{\ifcase#1 0\or1\or2\or3\or4\or5\or6\or7\or8\or9\or
 A\or B\or C\or D\or E\or F\fi }

\makeatletter
\ifx\CUP@mtlplain@loaded\undefined
\else
\fi
\makeatother
 \makeatletter
 \ifx\CUP@mtlplain@loaded\undefined
   \font\tenbmi=cmmib10 at 10pt
   \font\sevenbmi=cmmib10 at 7pt
   \font\fivebmi=cmmib10 at 5pt

   \newfam\bmifam
   \textfont\bmifam=\tenbmi
   \scriptfont\bmifam=\sevenbmi
   \scriptscriptfont\bmifam=\fivebmi
   
 \fi
 \makeatother
\ifnfsstwo

\fi
\ifnfssone

\fi
\ifoldfss

\fi

\mathchardef\varLambda="0103

\makeatletter
\ifx\CUP@mtlplain@loaded\undefined
\else
\fi
\makeatother
\makeatletter
\ifx\CUP@mtlplain@loaded\undefined
  \font\tenbms=cmbsy10
  \font\sevenbms=cmbsy10 at 7pt
  \font\fivebms=cmbsy10 at 5pt
  \newfam\bmsfam
  \textfont\bmsfam=\tenbms
  \scriptfont\bmsfam=\sevenbms
  \scriptscriptfont\bmsfam=\fivebms

  \edef\bsy@{\hexnumber\bmsfam}
  \mathchardef\bnabla="0\bsy@72
\fi
\makeatother
\begin{document}
\ifnfssone
\else
  \ifnfsstwo
  \else
    \ifoldfss
      \let\mathcal\cal
      \let\mathrm\rm
      \let\mathsf\sf
    \fi
  \fi
\fi


  \title[Conference Summary]{Conference Summary\\
    Supernovae and Gamma-Ray Bursts}

  \author[J. C. Wheeler]{%
  J.\ns
  C\ls R\ls A\ls I\ls G\ns
  W\ls H\ls E\ls E\ls L\ls E\ls R\ns}
  \affiliation{Department of Astronomy,
    University of Texas, Austin, TX , 78712, USA}
  \maketitle

\begin{abstract}
There are hints that nearby Type~Ia supernovae may be a little
different than those at large redshift.  Confidence in the conclusion that
there is a cosmological constant and an accelerating Universe
thus still requires the hard work of sorting out potential systematic
effects. Polarization data show that
core-collapse supernovae (Type II and Ib/c) probably depart strongly
from spherical symmetry.
Evidence for exceedingly energetic supernovae must be considered
self-consistently with evidence that they are asymmetric,
a condition that affects energy estimates.
Jets arising near the compact object can produce such
asymmetries.  There is growing conviction that
gamma-ray bursts intrinsically involve collimated or jet-like flow
and hence that they are also strongly asymmetric.
SN~1998bw is a potential rosetta stone that will
help to sort out the physics of explosive events.  Are 
events like SN~1998bw more closely related to ``ordinary" supernovae or
``hypernovae?"  Do they leave behind
neutron stars as ``ordinary" pulsars or ``magnetars" or is the
remnant a black hole?  Are any of these events associated
with classic cosmic gamma-ray bursts as suggested by the supernova-like
modulation of the afterglows of GRB~970228, GRB~980326 and GRB~990712?
\end{abstract}

New data is driving both supernova and \grb\ 
research and suggesting that these subjects may
be related.  A central issue in both areas is the breaking of
spherical symmetry.
This conference celebrated as much as any single thing the emergence
of evidence and argument for strong breakdown of spherical
symmetry for both supernovae and \grbs\ and especially
the resulting potential for links between them, whether those links are
supernovae or ``hypernovae."
Supernova studies have also revolutionized the study of cosmology.
There spherical symmetry is not at issue, but the prospect of an
accelerating Universe presents many challenges.

I cannot synthesize all the energetic work presented in oral talks
and poster presentations, never mind over coffee and dinner, at
this symposium that mixed two ``exploding" fields.  I will attempt
to give a summary of certain highlights and connective themes
that I think will set the course for future research.  In \S 2,
I give a brief summary of the exciting work on SN~Ia, their
associated physics, and their application to cosmology.
Some perspectives on \grbs\ and their possible link
to supernovae or ``hypernovae" are presented in \S 3. New
results on the propagation of a jet through a stellar core
are given in \S 4.  Some perspectives and conclusions are
presented in \S 5.

\section{Type Ia Supernovae and Cosmology}

There is a general concensus that the strong majority, if not all,
Type Ia supernovae arise in carbon/oxygen white dwarfs of very
near the Chandrasekhar mass.  The evidence in favor of this was
given by Livio (2000) who also summarized the problems of
understanding the binary evolution that allows a sufficient number of
white dwarfs to grow to carbon ignition at the mass limit.  Most of
the observed Type Ia are ``Branch normal," (Branch, Fisher \& Nugent, 1993)
and allowance for
deviations from ``standard candles" can be made rather successfully
with one-parameter brightness/decline rate relations (Phillips 1993;
Riess, Press \& Kirshner 1996; Perlmutter et al. 1999; Sandage 2000).

A dichotomy of thinking arises at this point.  The theorists say
that a one-parameter brightness/decline rate relation cannot
be the whole story.  Theory suggests appreciable variation with
input parameters (Khokhlov 2000; H\"oflich \& Dominguez 2000)
and the observations themselves suggest departures
from one-parameter relations.  Observers, on the other hand,
point out that utilizing these one-parameter relations
works remarkably well in practice in correcting for deviations
(Schmidt 2000; Perlmutter 2000; Sandage 2000).
The conclusion, as emphasized by H\"oflich \& Dominguez (2000) seems to be
that some of the input parameters that are varied independently
in the evolutionary and dynamic models -- the carbon
ignition density, the density of transition from subsonic
deflagration to supersonic detonation (Khokhlov, Oran \& Wheeler
1997a,b; Niemeyer \& Woosley 1997; 
Khokhlov 2000), rotation, progenitor mass, progenitor metallicity --
must be correlated in ways we have yet to elucidate.

To make progress, we must understand the combustion physics
(Niemeyer 1999; Khokh-lov 2000),
and we need to better understand the progenitor evolution (Livio 2000;
H\"oflich \& Dominguez 2000).  I would be delighted to witness even a shred of
direct observational evidence that Type Ia are in binary systems, a
conclusion to which I hold firmly despite vivid understanding that there
is no proof.

Despite the uncertainties that still plague work on SN~Ia, they
have been used with great effect to explore cosmological issues.
Theory has been used to compare models with individual supernovae
in a way that does not require secondary distance calibration.
The result is that the value of the Hubble constant is estimated
to be 67$\pm$9 km s$^{-1}$ Mpc$^{-1}$ (H\"oflich \& Khokhlov 1996).
Supernova observations calibrated with Cepheid variables give
values in the range 60$\pm$9 km s$^{-1}$ Mpc$^{-1}$ (Sandage 2000)
to 65.2$\pm$1.3 km s$^{-1}$ Mpc$^{-1}$ (Riess et al. 1998).

The application to cosmology has been even more startling (Riess
et al. 1998; Perlmutter et al. 1999; Perlmutter 2000; Schmidt 2000).
The value of the normalized cosmological matter density
derived from Type Ia supernovae is
$\Omega_m~\sim~1/3$ and that of the cosmological constant,
$\Omega_{\Lambda}~\sim2/3$.  This raises two sets of issues.
One is a possible new view of the Universe.  With a cosmological
constant that might not be ``constant," there is the potential
for closed Universes that expand forever or open Universes that
collapse.  There must be a consideration of new physics to
understand the microscopic origin of the vacuum energy that
poses as a cosmological constant and why it is so small,
but not zero, at just this epoch.

The other issue is, to use
Brian Schmidt's (2000) phrase, the ``mundane."  There may be
subtle systematic effects that bias the estimates of brightness
of supernovae as a function of redshift and which masquerade
as the effect of a cosmological constant.  Howell, Wang \& Wheeler
(1999) have noted that the nearby Type Ia that are
used to calibrate the light curve decline relation are primarily
discovered photographically so that they are susceptible to the
``Shaw effect" of being lost near the centers of galaxies due
to saturation,
whereas the deep searches are done with CCDs that
are less susceptible to this effect.  The properties of
the Type Ia's vary with galactic radius (Wang, H\"oflich \& Wheeler 1997;
Riess et al. 1999a) and the radial distributions of the
calibration sample and the distant sample are distinctly different.
This difference may be removed by light curve decline corrections,
but such systematic effects need more study.  Riess et al (1999b,c)
report that nearby Type Ia might have systematically slower
rise times than the cosmological events for similar decay times.
Suntzeff (1999) notes that the sample of events
that show distinct departures from a one-parameter
brightness/decline relation is real and growing
and that while 6 of 40 Type Ia in a nearby sample are of the very bright
kind, no events like SN~1991T have been observed in the
much larger deep sample.

All these developments are hints of systematic effects that must be better
understood, both physically and observationally.
Resolving this issue of the cosmological versus the mundane
one way or the other will take several year's hard, slogging
work by both theorists and observers in the supernova community.
The necessary program of careful comparison of the spectra
and light curves of near, intermediate, and far supernovae is underway.

\section{Gamma-Ray Bursts, Hypernovae, and Supernovae}

\subsection{Gamma-Ray Bursts, Collimation, and Jets}

The past year has seen discussions of extreme energies, ranging
up to $3\times10^{54}$ ergs (Kulkarni et al. 1999), and extreme
degrees of collimation (Wang \& Wheeler 1998).  The community seems
to have gotten those excesses out of its system and is now
buckling down to the hard work of figuring out the true nature of
the \grbs\ and their afterglows.  Table 1 gives a list of the
\grbs\ with observed X-ray afterglows.  Table 2 gives a compilation of
some of the relevant properties of \grbs\ in the afterglow era
(see also http://astro.uchicago.edu/home/web/reichart/grb/grb.html;
http://www.aip.de/People/JGreiner/grbgen.html).  Excellent 
reviews of \grbs\ and afterglows were given by Paczy\'nski (2000), 
Fishman (2000), Piro (2000), Fruchter (2000), Kulkarni (2000), Rees (2000),  
and Piran (2000).

\setcounter{table}{0}

\begin{table}

\caption{Gamma-Ray Bursts with X-ray afterglows$^1$}

\label{wheeler-tab1}

\begin{center}

\scriptsize

\begin{tabular}{llllll}\\\hline\hline

970111 &970828 &980326? &980613 &990123  &990704\\

970228 &971024 &980329 &980703  &990506  &990705\\

970402 &971214 &980425 &980923  &990510  &990712\\

970508 &971227? &980519 &981226 &990520 &990806\\

970815 &        &       &        &990627 &\\\hline

\end{tabular}

\end{center}

\noindent
$^1$ http://www.aip.de/People/JGreiner/grbgen.html

\end{table}

\begin{figure}
\setcounter{figure}{0}
\setcounter{table}{1}
\psfig{figure=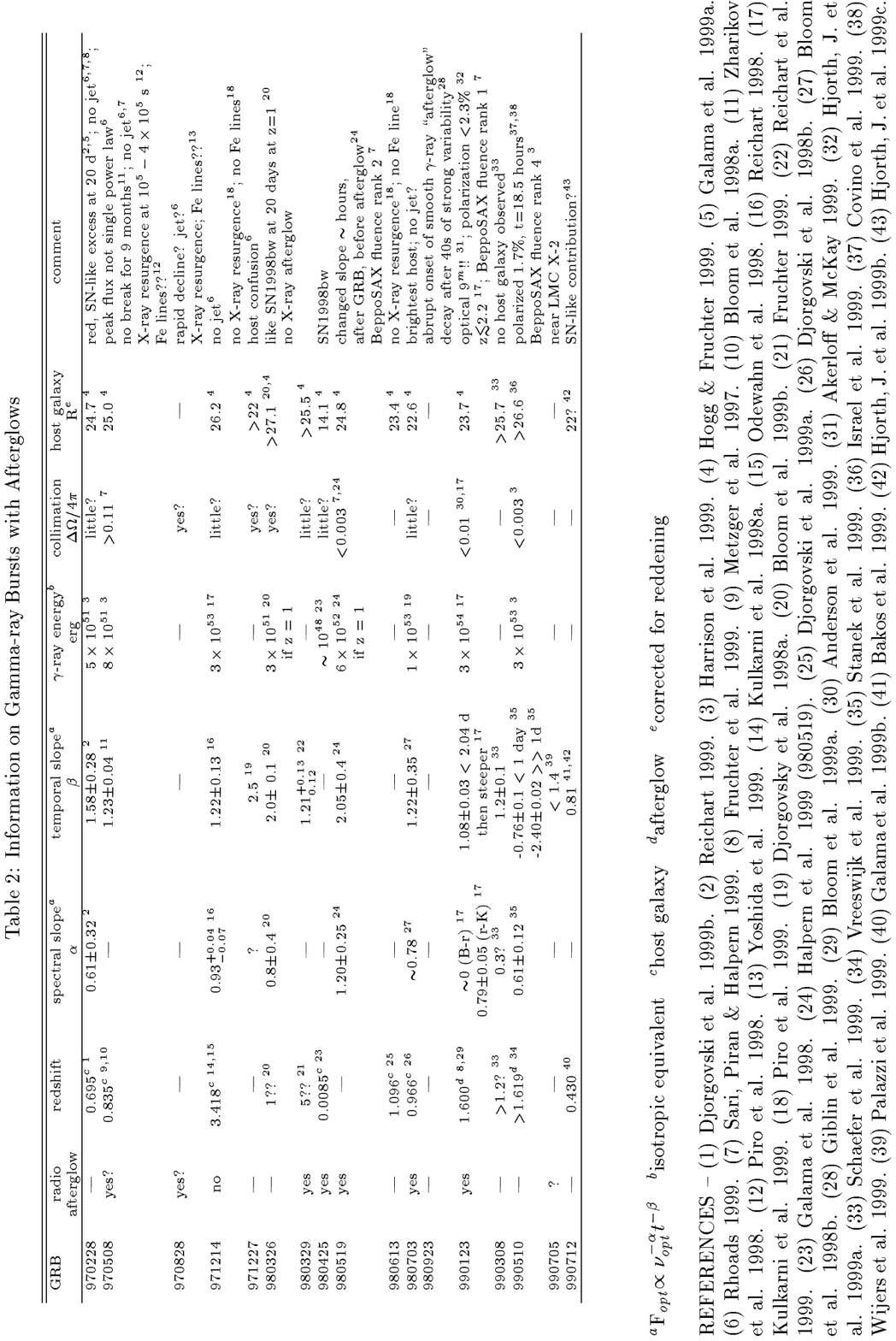,angle=180}
\label{wheeler-tab2}
\end{figure}

Prior to BeppoSAX and the discovery of the afterglows, there was
some discussion in the literature of the possibility of
collimation (Woosley 1993; Rhoads 1997), but by and
large the general community paid only lip service to collimation,
if any mention was made of it at all.  In the last year, the
phrase ``isotropic equivalent" has become a common and even
mandatory part of the vocabulary of papers on \grbs\ and afterglows.  Even
more recently, judging by postings to ``astro-ph," the specific
phrase ``jets" has become common parlance.  Over this year
there has been a maturity from musing about collimation to wide-spread
general acceptance that collimation is a critical aspect of some, if
not all, \grbs.

Judging from presentations at this meeting, I have already lost this
battle, but I would like to plead with the community to use
the words ``collimation" or ``jets" or their equivalent when non-spherical
flow is implied rather than the phrase ``beaming."  The latter is
often clear in context, but is prone to confusion with the Lorentz
beaming that is purely a kinematic effect of relativistic motion.
Authors who want to discuss both collimation and Lorentz beaming in
the same paper are inviting confusion if they do not clearly
discriminate.  The total energy can be determined straightforwardly
in principle by multiplying the isotropic equivalent energy with
the collimation factor, $\Delta~\Omega/4\pi$, but the discussion of
luminosity that involves Doppler factors in a more complex way can be
hopelessly muddled if the effects of collimation and Lorentz beaming
are not clearly dilineated.

At this conference, we heard about energies associated with
\grbs\ ranging from $4\times10^{50}$ to $3\times10^{54}$~ergs
(Frail 2000; Kulkarni 2000).  In fact, energies of
$4\times10^{50}$ to $10^{52}$ ergs were associated with
the same object!  Frail (2000) made a clear case for the
need to be careful in how energies are assigned to afterglows.
The energies are sensitive to model parameters (e.g break frequencies),
the estimates of which are in turn subject to observational
uncertainties.  The lower energies represent circumstantial evidence
that the highest energies are, indeed, only isotropic equivalents
and must be scaled down by a substantial collimation factor.

One expects the decay slopes to become steeper as a collimated
flow slows and spreads laterally (Rhoads 1997,1999),
and evidence for steeper slopes and even breaks in the
slope have now been observed (Kulkarni et al. 1999; Sari, Piran
\& Halpern 1999; Harrison et al. 1999; Table 2).
This evidence for collimation
is only circumstantial, and the rate of change of the slope
may be a gradual rather than sudden process (Panaitescu \&
M\'eszar\'os 1998; Moderski, Sikora \& Bulik 1999).
The slope can also be affected by density gradients in
the surrounding medium (Chevalier \& Li 1999) and by
variations in power output in the underlying ``machine"
(Li \& Chevalier 1999; Dai \& Lu 1999a,b).  With these caveats,
the strongest evidence for collimation is in the brightest sources, with
GRB~990123 with an isotropic equivalent
\gr\ energy of $3\times10^{54}$~ergs
having an estimated collimation factor of $\sim~0.01$ for
an actual \gr\ energy of $\sim~3\times10^{52}$~ergs (Kulkarni et al. 1999)
and GRB~980519 and GRB~990510 (which went off just after this conference)
having a beaming factor of less than $\sim0.003$
(Sari, Piran \& Halpern 1999; Harrison et al. 1999).
This suggests that three of the \grbs\ with the highest isotropic
equivalent energy are also the ones with strongest evidence for
collimation.

Another recent argument in favor of some form of collimation
comes from observations of polarization of \grbs.  The synchrotron
radiation from \grbs\ and their afterglows should be strongly
polarized with the degree of polarization from a single patch
ranging as high as 70 \% (Sari 1999), but if the source is
spherically symmetric and the field tangled, then the degree
of polarization will be reduced.  In this context, the
upper limit to the polarization of 2.3\% for GRB~990123
(Hjorth et al. 1999a) and the measured polarization
of 1.7 \% in GRB~990510 (Covino et al. 1999; Wijers et al. 1999)
are very significant for being so small. This may suggest that
in addition to a tangled magnetic field, we are observing
collimated flow and emission
(Gruzinov 1999;  Ghisellini \& Lazzati 1999; Sari 1999).
Note that, as Gruzinov stresses, the origin of the magnetic
field, assumed to be in near equipartition, remains a major stumbling block
for the relativistic blast wave synchrotron emission model.

With the evidence for collimation,
there is a suggestion that the true \gr\ energies of all
the \grbs\ are in the range $10^{50}$ - $10^{52}$ ergs, from
0.1 to 10 times the canonical energy of a supernova, but still
substantially less than the binding energy of a single neutron star.
It is also true that as the evidence for collimation has grown,
so have the highest recorded isotropic equivalent energies,
leaving most people with the feeling that, at best, the target
energy is near the upper end of this range, $\sim10^{52}$ ergs.
The issue becomes the nature of supernovae, hypernovae, and their
link to \grbs.

\subsection{Polarization of Supernovae}

Like many people in the supernova community, we at the University
of Texas got actively involved in the supernova/soft-gamma-ray
repeater/magnetar/\grb\ topic with the advent of SN~1998bw and
its possible connection to GRB~980425.  We brought a different
perspective to this issue because of work we have done over
the last four years on supernova spectropolarimetry.

We have been making spectropolarimetric observations of all
accessible supernovae at McDonald Observatory (Wang et al. 1996;
Wheeler, Wang \& H\"oflich 1999; Wheeler, H\"oflich \& Wang 1999).
The result has been that most Type Ia have low polarization and
hence are substantially spherically symmetric.  Many have only
upper limits of order 0.1 - 0.2\%.  A few have detected, but
low polarization, of order 0.2\%.  The polarization observed
is consistent with theoretical models of delayed detonation
models (Wang, Wheeler \& H\"oflich 1997) and may be a useful
probe of the combustion physics.  We have detected one exception,
SN~1997bp, which was observed a week before maximum light to
have a polarization of about 1\%.  The polarization was low in
post-maximum spectra, but this event remains a challenge to
understand.  It is important to establish whether such events are
common, the physical reason for the large polarization, and
whether or not there could be an asymmetric luminosity distribution
that could affect estimates of cosmological parameters.

More importantly in the current context are our observations
of presumed core-collapse events, Type II and Type Ib/c.  We have
found that all such events are polarized at about the 1\% level.
So far there have been no exceptions in about a dozen events.
There could be a myriad reasons for polarization, but our data
suggest a very important trend: the smaller the hydrogen envelope,
the larger the observed polarization.  As an example, SN~1987A
with a 10\m\ envelope had a polarization of about 0.5\% ( M\'endez et al.
1988), SN~1993J and a very similar object, SN~1996cb,
with small hydrogen envelopes, $\sim0.1$\m, were polarized
at the 1-2\% level (Trammell, Hines \& Wheeler 1993;
Tran et al. 1997), and Type Ic SN~1997X which
showed no substantial hydrogen nor helium was polarized at perhaps
greater than 3\% (Wheeler, H\"oflich \& Wang 1999).
These are difficult observations requiring
special care in the reduction to remove the effects of the ISM
(the latter greatly aided by wavelength and temporal coverage),
and there is a pressing need to expand the statistical sample.
Nevertheless, this trend suggests that the core-collapse process
itself is strongly asymmetric and that evidence for that asymmetry
is damped by the addition of outer envelope material.

The level of polarization we have observed for core collapse events,
$\sim1$\%, requires a substantial asymmetry with axis ratios
of order 2 to 1 (H\"oflich 1995).
Asymmetric explosions tend to turn spherical as
they expand, so to leave an imprint of an asymmetry of this level
in the homologously expanding matter requires a substantially larger
asymmetric input of energy or momentum in the explosion process
itself.
These factors led us to the hypothesis that the core collapse
process is intrinsically strongly asymmetric, much more so than
current collapse calculations involving convectively unstable
neutron stars.  It was in this context that we greeted the
news of SN~1998bw a year ago.

\subsection{SN~1998bw and GRB~980425}

SN~1998bw was a shock to both the supernova and \grb\ communities.
For the supernova community, it was clearly an odd and exciting event,
even without the possible association with GRB~980425.  Although it
resembled a Type Ic in the sense that there was no obvious evidence for
H and He, it was different than the canonical Type Ic. It was
very bright, it showed very large velocities, and it had a
very bright radio source.  While not the brightest supernova radio event on
record (Weiler 2000), the radio source associated with SN~1998bw
was undoubtedly very luminous and arguments based on the brightness
temperature alone suggested relativistic motion (Kulkarni et al. 1998b)
depending on whether or not one assumed magnetic field
equipartition (Waxman \& Loeb 1999).  On the other hand, the \gr\
community had just gone through the catharsis of proof that
\grbs\ were, indeed, at cosmological distances as strongly suggested
by the isotropy of the BATSE sources (Fishman 1995, 2000) and confirmed
by redshifts measured for a number of the events with detected
afterglows.  The identification of SN~1998bw with GRB~980425 immediately
confused the issue by raising the prospect of substantially
different sources of \grbs\ that could not be easily differentiated
by their \gr\ flux, fluence, time history, or spectra.

I took the opportunity of this meeting to test a rift I had suspected
since the Texas Symposium in Paris in December, 1999.  I inquired
as to the people in the audience who primarily identified themselves
with the supernova community and those who identified themselves
primarily with the \grb\ community.  I then asked for a show of hands
of those who had reservations about the identification of SN~1998bw
with GRB~980425 and of those who were convinced that the two objects
were one and the same.  While there were some cross-over votes,
this excercise basically confirmed my suspicion.  The supernova
community has bought this identification hook, line, and sinker,
based on the odd properties of the supernova.  The \grb\ community
remains substantially suspicious despite the low {\it a priori}
probability, $\sim 10^{-4}$ (Galama et al. 1998), of an
accidental alignment.  One does not do science by democratic vote,
and this is more an excercise of sociology than science, so I
leave the reader to contemplate the meaning, if any, of
the result (and to criticize the unscientific sampling method).

The issues raised by SN~1998bw are these.  If one believes that
SN~1998bw produced GRB~980425, then there are more than one
type of \grb\ that cannot be easily distinguished by \gr\
properties.  If one does not believe that these two events
are identical, there is still a weird supernova to explain
and the standard mystery of the nature of the cosmic \grbs\
remains.  All these possibilities bring with them the
strong suggestion that, for both the supernova and for
\grbs, asymmetries and collimation are the rule, not the exception.

Even if one believes the identification of
SN~1998bw with the \grb\ there are substantially different
interpretations of the event.  Was it a version of an
``ordinary" supernova or was it something better labeled
a ``hypernova?"  The phrase ``hypernova" was coined in this
context by Paczy\'nski (1998, 2000), who meant by it any
event that produced an isotropic equivalent luminosity
substantially larger than a supernova, whatever the source of
that luminosity, core collapse or merging compact stars.
The phrase ``hypernova" has taken on a somewhat more specific
meaning in the supernova community in the context of attempts
to understand the nature of SN~1998bw and possibly related
events.

The reason for this evolution of the definition of ``hypernova"
is that attempts to construct models for SN~1998bw
have led to the suggestion that the energy must be substantially
in excess of $10^{52}$ ergs (Iwamoto et al. 1998; Woosley,
Eastman \& Schmidt 1999; Branch 2000; Nomoto 2000; Woosley 2000)
and perhaps also that it had very large ejecta mass and radioactive
nickel mass, the latter in excess of 0.5 \m\ (Iwamoto et al. 1998; Woosley,
Eastman \& Schmidt 1999).  These models were all, by assumption,
spherically symmetric, and, if nothing else, ignore the
observations that SN~1998bw was polarized.  On the other hand, H\"oflich,
Wheeler \& Wang (1999) have argued that a sufficiently asymmetric
model to account for the polarization (an ellipsoid of axis
ratio of 2 to 1) could account for the bolometric and
multi-color light curves near maximum light.  With the
proper viewing angle, within about 35${^{\rm o}}$  of the symmetry axis
for a model with oblate isodensity contours, the luminosity
could be substantially higher than the mean spherical equivalent
and the peak of the light curve reproduced in a model with
an ejecta kinetic energy of $2\times10^{51}$ ergs, 2\m\ of ejecta,
and 0.2\m\ of \ni.  The kinetic energy of this model is a little
higher than average, but well in the range of energies deduced for
standard core-collapse events, and the other parameters are quite nominal.
SN~1998bw may have been a ``hypernova" requiring
expansion energy more than 10 times that normally associated
with supernovae, but that is certainly not necessarily so.

Given its important role, SN~1998bw received a great deal of
attention at the conference.  It is poignant to note at a
conference hosted by STScI that the request by the author
and Lifan Wang for Director's Discretionary Time was turned
down in the press of other observational priorities.  The
result is that there were no UV observations of SN~1998bw.  This lack
becomes more important as one ponders similar events at
large redshift where the UV spectrum is shifted to the visible.
We can guess what SN~1998bw would have revealed in such
observations, but we will never know.

One of the principle controversies over the identification of
SN~1998bw with GRB~980425 is the nature of the afterglow or
the lack thereof.  This issue was addressed anew by Pian (2000).
The first narrow-field instrument BeppoSAX observations
revealed two X-ray sources in the original wide-field detection
image. Neither corresponded to the location of the supernova.
One source (Source 1) was at first thought to be constant
and the other (Source 2) to decay in a few days in a manner consistent
with other observed X-ray afterglows.  This observation has been
the origin of understandable suspicion by many that the association
of the supernova and the \grb\ were accidental despite the
low probability.

Recalibration of positions (Pian et al. 1998) revealed
that Source 1 was coincident within the errors with SN~1998bw,
but that Source 2 was definitely not associated
with the supernova.  Further observations showed that Source
1 did vary, but slowly, perhaps a factor of 2 in 10 days.
At this conference, Pian presented evidence that Source 2
also declined slowly or was even substantially constant over
an interval of more than 100 days.  Taken at face value, the
data presented by Pian suggest that both Source 1 and Source 2
declined rather slowly after the first NFI observations beginning
about 1 day after detection.  The flux for both is about $10^5$
less than that first detected in the WFC.  It is possible
that neither Source 1 nor Source 2 represent an afterglow.  It
is possible that the afterglow decayed so rapidly from the
WFC detection that only background sources were detected at
Source 1 and Source 2.  The interpretation of the data depend
substantially on the confidence placed in the detection of
Source 2 at the 3$\sigma$ level after day 100. If this detection
is true, then it seems unlikely that either Source 1 or Source 2
are an afterglow.  If this detection is only an upper limit,
then it is still conceivable that Source 2 is an afterglow and
the identification of GRB~980425 with SN~1998bw is still open
to question.  Observations of these sources by ASCA in
April may have yielded only upper limits (Harrison 1999), but
these may help to resolve this issue.

Danziger et al. (2000) presented photometric, spectroscopic
and spectropolarimetric observations of SN~1998bw from ESO.
Danziger concluded that
the object was asymmetric, in substantial agreement with
H\"oflich, Wheeler \& Wang (1999).  Nomoto (2000) presented
model light curve calculations.  He noted that the spherically
symmetric models required ``hypernova" energies to fit the
peak, but that the same models declined too rapidly to
fit the tail.  He pointed out that one way to account for
the tail was to invoke a smaller expansion energy to get
greater trapping of \grs\ at later times.  He concluded
that this implicit contradiction was evidence for
asymmetry.  This underlines the basic point of H\"oflich,
Wheeler \& Wang (1999) that the energy cannot be determined
independently of considerations of asymmetry for the dynamics
and radiative transfer.

Danziger et al. (2000) also presented nebular spectra of SN~1998bw.
These spectra revealed futher peculiarities.  The line of
[OI] $\lambda\lambda$ 6300,6364 was broader than the lines
of [Fe II].   The Fe lines were comparable in width to
those of normal Type Ic in contrast to the expectation from
the basic hypernova models that, with their very high
energies, have very high velocities, $\gta 4000$ \kms, in
the inner, iron-rich, regions.  Discussion with Nomoto and Mazzali
suggested that the spectra could be fit with models,
but only by adding {\it ad hoc} inner regions of slower
moving matter.  Such slow matter may be produced in a
more realistic, multi-dimensional model, but it is not
predicted in the spherically symmetric ``hypernova" models.

Branch (2000) presented simple
atmosphere models that illustrated the systematic differences
between SN~1994I, SN~1997ef, and SN~1998bw.  SN~1994I was
a canonical, well-studied Type Ic supernova.  SN~1997ef was
also labeled a Type Ic, but while showing a normal peak
luminosity, it had higher velocity at the photosphere than
SN~1994I.  Branch illustrated how the increased broadening
of the lines carved out the red continuum and led to a
rather steep decline from the blue.  With the even higher
photospheric velocities of SN~1998bw, Branch provided convincing
evidence that the unprecedentedly steep decline from about
4000 to 5000 Angstroms in the continuum of SN~1998bw near maximum
light could be explained.  Coupling his atmosphere models
with the assumption of spherical symmetry, Branch made
estimates of the kinetic energy of each event, concluding that
SN~1994I was consistent with $\sim10^{51}$ ergs, but that
both SN~1997ef and SN~1998bw could require ``hypernovae"
energies of $\gta 10^{52}$ ergs.  Branch pointed out
that the ball was in the court of advocates of asymmetric
models to show that these observations could be explained
self-consistently with asymmetries and modest energy.
That is completely correct.

An excellent light curve of SN~1998bw is presented by
McKenzie \& Schaeffer (1999).  They have an well
time-sampled data set that shows that after an early steep decline
from maximum for about 25 days, B, V, and I have declined
in a precisely exponential manner.  The slopes in the
three bands are slightly different and all three are
steeper than that expected for \co\ decay and full trapping
of \grs.  McKenzie \& Schaeffer note that because there
must be some leakage of \grs\ they can only set a lower
limit to the amount of \ni\ produced in SN~1998bw which
they determine to be 0.22$\pm$0.09 \m.  This lower limit
is close to that estimated for the asymmetric models of
H\"oflich, Wheeler \& Wang (1999) and substantially less
than the hypernova models of Iwamoto et al. (1998)
and Woosley, Eastman \& Schmidt (1999).  Since the light curves
are so precisely exponential, the \gr\ trapping fraction
cannot be changing substantially.  This suggests that
while less than unity, the trapping fraction is substantially
greater than 50\% or the light curves would be steeply
declining to the limit set by positron trapping.  This
argument suggests that the lower limit set by McKenzie
\& Schaeffer may be near the actual nickel mass produced
by SN~1998bw and in contradiction to the hypernova models.

\subsection{Other Possible Supernova/Gamma-Ray Burst Connections}

SN~1998bw/GRB~980425 is the most famous and best
established supernova/\grb\ connection (despite or because
of its debated reality), but other arguments have accumulated
for such a connection.  Some candidates as of this writing
are given in Table 3.


\setcounter{table}{2}

\begin{table}

\caption{SN/GRB Candidates}

\label{wheeler-tab3}

\begin{center}

\scriptsize

\begin{tabular}{llcccccl}\\\hline\hline

SN &GRB &\multicolumn{2}{c}{SN properties}
&\multicolumn{3}{c}{GRB properties} &REFERENCES\\

& &spectra &light curve &flux$^a$ &fluence$^b$ &duration(s) &\\\hline

~~---  &970228$^{c}$&$\sim$98bw? &$\sim$98bw? &---  &---              &80
&1, 2\\
1997cy &970514?     &IIn         &            &     &$4\times10^{-7}$ &1.3 &3\\
1997ef & 970125??   &$\sim$98bw? &            &     &                 &    &4\\
       &971115??    &            &            &     &                 &    &\\
~~---  &980326      &$\sim$98bw? &$\sim$98bw?
&$8\times10^{-7}$&$7\times10^{-7}$ &0.2 & 5, 6\\
1998bw &980425      &$\sim$Ic    &            &     &$4\times10^{-6}$ &35  &7\\
1999E  &980910?     &$\sim$97cy  &M$_V<$-19.4 &     &$2\times10^{-7}$ &---
&8, 9, 10, 11\\
~~---  &990712      &$\sim$98bw? &$\sim$98bw? &---  &---              & ---
&12 \\\hline                                                                      
\end{tabular}

\end{center}

\noindent
$^a$erg cm$^{-2}$ s$^{-1}$~~$^b$erg cm$^{-2}$~~$^c$ BATSE behind Earth

\medskip
\noindent
REFERENCES - (1) Reichart (1999). (2) Galama et al. (1999).
(3) Germany et al. (1999). (4) Wang \& Wheeler (1998).
(5) Bloom et al. (1999). (6) Briggs et al. (1998).
(7) Galama et al. (1998). (8) Filippenko, Leonard \& Riess (1999).
(9) Jha et al. (1999). (10) Capellaro, Turatto \& Mazzali (1999).
(11) Thorsett \& Hogg (1999). (12) Hjorth et al. (1999c).

\end{table}

Germany et al. have discussed the case of SN~1997cy.  This
supernova was odd in its own way and unlike SN~1998bw in
many substantial ways. The supernova occurred in a low
surface brightness galaxy at a redshift of z = 0.063.  The
date of the explosion is uncertain by a few months.
The spectrum is characterized by a very strong line of
H$\alpha$, unlike SN~1998bw which showed no evidence for
hydrogen.  The H$\alpha$ line showed both broad and
narrow components reminiscent of Type IIn supernovae
(Schlegel 1990).   SN~1997cy also showed lines of Fe II
and [Fe III] that are more characteristic of the
nebular phase of Type Ia events.
The light curve of SN~1997cy followed the decay slope
of $^{56}$Co for about 60 days after discovery.  The
light curve then flattened for 200 days and then
proceeded to a more steep decline.  Assuming the early
part of the light curve to be due to the trapping of
\grs\ from $^{56}$Co decay, Germany et al. deduce that
SN~1997cy ejected 2 \m\ of $^{56}$Ni.  They tentatively
ascribe the subsequent flattening and decline of the light
curve to circumstellar interaction.  Germany et al. note
that SN~1997cy was about a factor of 50\% brighter at
discovery than SN~1998bw was at maximum. Late-time radio
observations, 16 months after the explosion, revealed no
detectable source.   Germany et al.
argue for a possible connection of SN~1997cy with GRB~970514.
This burst lasted $\lta$ 1 s and was classified as a
high-energy event.   They estimate the chance association
of the two events to be about 1\%.  If the events were associated,
then the \gr\ energy in the burst was about $4\times10^{48}$ ergs,
comparable to, but somewhat larger than, that ascribed to
SN~1998bw/GRB~980425.  Germany et al. also note that the
decay of GRB~970508 was similar in slope to SN~1997cy and
$^{56}$Co.  They also note that SN~1999E had a spectrum
similar to SN~1997cy, that SN~1999E was especially bright,
and that it might be temporally linked to GRB~980910.
One must be somewhat cautious in interpreting the data from
SN~1997cy, especially the key observation that the light curve
traced cobalt decay.  There is no question that the supernova
was bright at its redshift.  On the other hand, the light
curve was observed to fall at the rate of $^{56}$Co for
only about 60 days, barely half a $^{56}$Co e-fold time. 
If the association with GRB~970514 is questioned, then the
time of explosion is uncertain and this also impacts the 
amount of $^{56}$Co one would attribute to the event, even
accepting that $^{56}$Co decay is observed.

Another class of association of \grbs\ and supernovae
comes from the discovery of transient brightening or
modulation of afterglows.  Bloom et al. (1999b)
argue that a brightening of the light curve of GRB~980326
can be interpreted as the addition of the light of an
event like SN~1998bw about 20 days after the explosion
if the supernova were at about a  redshift of 1.  The
optical transient became about 60 times brighter than
expected from an extrapolation of the decline of flux at
earlier times.  In addition, Keck spectra showed that
the continuum spectrum changed from being blue to being
red.  The latter is roughly consistent with the radiation
from a supernova photosphere, but inconsistent with
synchrotron radiation as might have occurred if
there were delayed energy input or
the blast wave ran into a dense cloud, possible alternative
models (Panaitescu, M\'esz\'aros \& Rees 1998; Dai \& Lu 1998a,b;
Piro et al. 1999a).  Bloom et al. note that special
circumstances might be necessary to reveal such a late-time
rebrightening: a rapid afterglow decay, and a low surface-brightness
host galaxy.

Reichart (1999) has advanced similar arguments in a study of
an earlier event, GRB 970228.  While there is no spectroscopy,
Reichart notes that there is much more thorough photometry
for GRB~970228 as compared to GRB~980326 and that there is
a measured redshift of z = 0.695 (Djorgovsky et al.  1999).
GRB~970228 showed strong reddening with time, a characteristic not explained
by standard relativistic blast wave models.  Reichart
argues that the afterglow data is not consistent with a single power
law spectrum nor a single power law temporal decay, but that
it is consisent with the U-band light curve of SN~1998bw
appropriately redshifted to the frame of GRB~970228.  The
spectral energy distribution is also consistent
with the convolution of a SN~1998bw-like event and the
power-law decline of a relativistic blastwave.   Similar
conclusions have been reached by Galama et al. (1999).  Interestingly,
this manifestation of a supernova-like resurgence is
seen despite a relatively slow decline in the early
afterglow $\propto t^{-1.58}$.  The host galaxy was
relatively dim compared to the early afterglow and the
``supernova" contribution.

The most recent suggestion of such a supernova-like modulation
has been given by Hjorth et al. (1999c) for GRB~990712.  They
argue that the R-band light curve is consistent with a 
temporal decay of the afterglow like $^{-1}$, a host galaxy
of R = 21.76, and a ``supernova" like SN~1998bw at the known
redshift of z = 0.430 (Galama et al. 1999b).

Although it is not at all clear that it should be presented
in this context, for completeness I will add the possibility
that SN~1987A produced a jet of some sort, if not a \grb.
Wang \& Wheeler (1998), Cen (1998), and Nakamura (1998) all
noted that if supernovae make jets that have some connection
to \grbs\ there might be some relevance to the ``mystery
spot" of SN~1987A.  Motivated by Cen's comments in this connection,
Nisenson \& Papaliolios (1999) re-examined their speckle data on
SN~1987A and argued in favor of both a jet and counter jet
from SN~1987A.  From the kinematics they concluded that the
counter jet must have moved at relativistic speeds.  Some
sort of jets might be produced in many core collapse events
(see \S3), but there is still debate and doubt on the
reality of these jets.  There is, of course, no direct link
to a \grb.  On the other hand, Nagataki (1999) makes a convincing
case that a jet-like explosion in SN~1987A can resolve many
of the issues of outward mixing of radioactive elements
and line profiles.

\section{Jet-Induced Supernovae}

The goal of producing robustly asymmetric supernovae, weak
\grbs\ of the sort observed in SN~1998bw/GRB~980425, and
perhaps collimated high-energy bursts of \grs\ that could
contribute to the cosmological \grbs\ suggests the following
general picture.  The best chance of imprinting the asymmetry
and producing some sort of \grb\ is in the absence of an
extended hydrogen envelope which could slow, delay, or disperse the
propagation of an asymmetric flow of energy from a newly-formed
neutron star.  This makes a Type Ib/c, and especially a Type Ic
configuration, a likely site for study, independent of the
similarities of SN~1998bw to Type Ic.

The progenitor of a Type Ic
is envisaged to be the core of a massive star, perhaps in excess
of 15\m\ on the main sequence, which has shed its hydrogen and
helium by a winds or binary mass transfer.  The iron core in
such a progenitor collapses to form a neutron star, and the
outer layers of Si, O, and C with longer free-fall times hover
momentarily.  The neutron star bounces and produces a standing
shock that stalls without inducing an explosion.  If the
neutron star is a pulsar, then it is possible to create an
MHD jet up the rotational axis at the time of the formation
of the neutron star as in the old calculation of Leblanc \& Wilson
(1970).  Such a jet could induce the requisite asymmetry in the
ejecta and accelerate in the density gradient to produce a
weak \grb.  If the pulsar were very highly magnetized, a
magnetar (Duncan \& Thompson 1992; Kouveliotou et al. 1998; Harding 2000),
then the pulsar could also potentially produce an intense
flux of Poynting radiation (Usov 1992,1994; Thompson 1994;
M\'esz\'aros \& Rees 1997; Blackman \& Yi 1998;
see also Ostriker \& Gunn 1971; Bisnovatyi-Kogan 1971).
This Poynting flux would tend to
flow out through the weakest part of the hovering mantle, namely
the wound punched by the MHD jet.  For very large magnetic fields,
the collimation and energetics could be sufficient to produce
a cosmic \grb.

This outline of a possible scenario illustrates
a host of places where more rigorous physics is needed to
determine whether the hypothesized outcome is reasonable.
One issue is the propagation of the initial MHD jet
out through the mantle and its effect on the star.  This
issue has been recently addressed by Khokhlov et al. (1999).

Khokhlov et al. adopt a progenitor Type Ib/c model consisting of a
a spherical helium star of radius $R_{\rm star} =  1.88 \times
10^{10}$~cm and mass $M_{\rm star}\simeq~4.1$\m.
The inner Fe/Si core with mass $M_{\rm core} \simeq 1.6
M_{\odot}$ and radius  $R_{\rm core} = 3.82 \times 10^8$~cm
is assumed to have collapsed on a timescale much faster
than the outer, lower-density material.  This core is replaced by a
point gravitational source with mass $ M_{\rm core}$ representing the
newly formed neutron star.  The remaining mass, $\simeq 2.5$\m,
consists of an O-Ne-Mg inner layer surrounded by
the C-O and He mantles.

The jets are assumed to enter the mantle at two polar
locations at $R_{\rm core}$.  An inflow with velocity $v_j$, density
$\rho_j$ and pressure $P_j$ is imposed.   The jet parameters are chosen to
represent the results of  LeBlanc \& Wilson (1970). At  $R_{\rm core}$,
the jet density and pressure are the same as those of the background
material, $\rho_j = 6.5 \times 10^5$~\gmcm3 and $P_j = 1.0 \times
10^{23}$~ergs cm$^{-3}$, respectively. The radii of the  cylindrical jets
entering the computational domain  are approximately $r_j = 1.2 \times
10^8$~cm.
The jet velocity at $R_{\rm core}$ was held
constant at $v_j = 3.22 \times 10^9$ \cms\ for 0.5 s.
This results in a mass flux rate of $\sim 9.5\times10^{31}$ \gm-s
with an energy deposition rate $dE / dt = 5\times 10^{50} $ \erg-s
for each jet.  After 0.5~s, the
velocity of the jets at $R_{\rm core}$ was gradually decreased to zero
at approximately 1~s. The total energy deposited by the jets was $E_j
\simeq 9\times 10^{50}$~ergs and the total mass ejected is $M_j \simeq
2\times 10^{32}$~grams or $\simeq 0.1$\m. These parameters are
consistent with, but somewhat less than, those of the LeBlanc-Wilson model.

As the jets
move outward, they remain collimated and do not  develop much internal
structure. A bow shock forms at the head of the jet and spreads in all
directions, roughly cylindrically around each jet.
The jet characteristic time,
$\tau_j\simeq 1$~s, is  much shorter than the sound crossing time of the
star, $\tau(R_{\rm star}) \simeq 10^3$~s.
The jets stay collimated enough to reach the surface as strong jets.
The stellar matter is shocked by the bow shock, and
acts as a high-pressure confining medium by forming a cocoon around
the jet. The sound crossing time of the dense O-Ne-Mg envelope,
$\tau(\sim 10^9~{\rm cm}) \simeq 10$~s, is only ten times longer than
$\tau_j$, and the jets are capable of penetrating this dense inner part
of the star in $\sim 2$~s.  By the time the jets penetrate into
the less dense C-O and He layers, the inflow of material into the jets has
been turned off. By this time, however, the jets have become long
bullets of high-density material moving through the background
low-density material almost ballistically. The higher pressures in
these jets cause them to spread laterally. This spreading is limited by
a secondary shock that forms around each jet between the jet and the
material already shocked by the bow shock.   The radius of the jets,
$\sim 3\times10^9$ cm as they emerge from the star, is larger than the
initial radius, $\sim 10^8$ cm,  but it is still significantly less
than the radius of the star.
After about 5.9 s, the bow shock reaches the edge of the star and
breaks through. Figure 1 shows the subsequent evolution of the star
after the breakthrough.  By $\simeq 20$~s, most of the material in the
jets has left the star and will propagate into the interstellar medium
ballistically.

\begin{figure}
\psfig{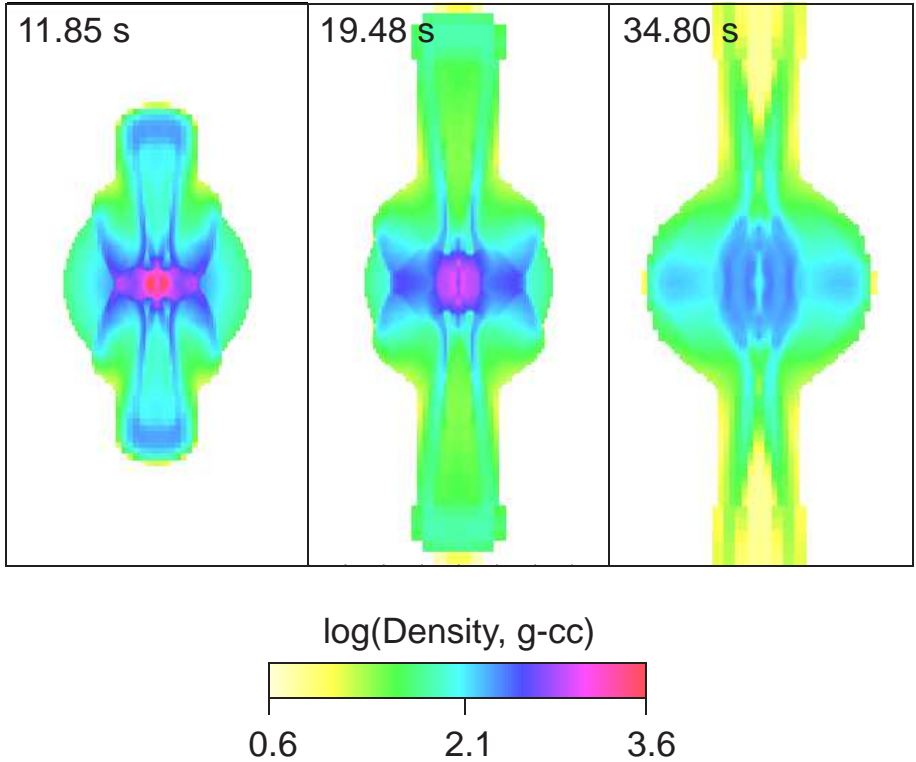}
\caption{
Jet evolution after breakout for the case of a progenitor which has
lost its hydrogen rich-envelope (from Khokhlov et al. 1999).
 The frames show the density in the x-z
plane passing through the center of the computational domain.
The time since the beginning of the simulation is given in the upper
left.
The sizes of the frames are $\Delta x =6.1\times 10^{10}$~cm and
$\Delta z= 1.125\times 10^{11}$~cm.
}
\label{figure 1}
\end{figure}

The laterally expanding bow shocks generated by the jets move
toward the equator where they collide with each other.
The result is that the material in the equatorial plane is
compressed and accelerated more than  material in other directions
(excluding the jet material). At $t\simeq 29$~s, the equatorial flow reaches
the outer edge of the star, and the star begins to settle into the free
expansion regime. The computation was terminated at $\simeq 35$~s,
before free expansion was attained. The stellar ejecta at this time is
highly asymmetric. The density contour of $50$ \gmcm3, which is
the average density of the ejecta at this time, forms an oblate
configuration with the equator-to-polar velocity and density ratios
$\simeq 2/1$ and $4/1$, respectively. Complex shock and rarefaction
interactions inside the expanding envelope will continue to change the
distribution of the parameters inside the ejecta. Nonetheless, we
expect that the resulting configuration will resemble an oblate
ellipsoid with a very high degree of asymmetry $\ge 2$.

The asymmetric explosion generated in this calculation provides
ejection velocities that are comparable to those observed in supernovae,
but with especially high velocities near the jet axis.
For this particular calculation,  an energy of
$9\times 10^{50}$ ergs is input at the base of the jets. This
energy is divided roughly equally between the emerging jets and
the bulk of the asymmetric ejecta.
The total mass in the two jets is $M_j
\simeq 0.1 M_\odot$ and the total kinetic energy is $E_j \simeq 5 \times
10^{50}$~ergs. The average velocity of the jets is about 25,000~ \kms.
The outer 2.5 \m\ of mantle material is ejected with kinetic
energy of $5\times 10^{50}$ ergs and average velocity $ 3,000 -
4,000$~ \kms.

The jet-induced explosion is entirely due to
the action of the jet on the surrounding star.
The mechanism that determines the energy of such an explosion is
related to the shut-off of the accretion onto the neutron star
by the lateral shocks that accelerate the material outwards.
The explosion thus does not depend
on neutrino transport or re-acceleration of the stalled shock.

The result of this calculation is a highly nonspherical
supernova explosion with two high-velocity jets of material moving in
polar directions and oblate, highly distorted ejecta containing
most of the supernova material.
This jet-induced explosion thus provides a satisfactory account of
the degree of polarization and asymmetry observed for typical
core-collapse supernovae.  The luminosity and photospheric
velocities will be a function of the aspect angle.  This
model gives at least the possibility of reproducing a standard
Type Ic like SN~1994I by observations along the equator where
the velocity and luminosity (and perhaps the \gr\ flux) will be minimum,
but the polarization will be maximum, of reproducing SN~1997ef
by an observation at intermediate angles, and reproducing
SN~1998bw by observation near the poles (within 35 degrees; H\"oflich,
Wheeler \& Wang 1999) where the velocity and luminosity will
be maximum and the polarization a minimum, albeit with perhaps
somewhat higher explosion energy.  This scheme might thus
account for many of the properties of these three events as outlined
by Branch (2000) without requiring a ``hypernova" for any of them.

The jets provide a large kinetic energy per unit solid angle.
When the jets break through the stellar photosphere, a small amount of
mass will be accelerated down the density gradient
to high velocities.  Khokhlov et al. (1999)
did not have sufficient resolution
to make quantitative predictions; however, a small fraction of the material
at the stellar surface had a velocity of up to $\sim$ 90,000 \kms.
There is thus a good liklihood of producing a weak \grb\
and a radio outburst of the
type seen in SN~1998bw/GRB~980425 by the relativistic shock
ejection mechanism of Colgate (1975).

Khokhlov et al. (1999)
assumed that the jets were generated by a magneto-rotational
mechanism during core collapse and neutron star formation (LeBlanc \&
Wilson 1970).  The LeBlanc \& Wilson 
calculation was criticized by Meier et al. (1976)
as requiring extreme parameters of the progenitor star.  These issues
need to be re-examined in the current context, but several
things are worth noting about the Meier et al. analysis.
They argue that the MHD
axial flow found by Leblanc \& Wilson will not propagate to the stellar
surface as a jet.  The calculation of Khokhlov et al.
shows this to be incorrect.
Meier et al. based their analysis on stellar evolution calculations of
the day, but they adopted a stellar core with central density of
about $10^{10}$~ \gmcm3 giving a binding energy of about $10^{52}$ ergs.
This exaggerates the binding energy of the initial core by about a
factor of 10 compared to modern calculations and gives an incorrectly
small value of a key parameter of Meier et al., the ratio of the
binding energy of the newly-formed neutron star to that of the
initial core.  Meier et al also did not consider the possibility of
an $\alpha-\Omega$ dynamo that could lead to exponential growth
of the magnetic field (Duncan \& Thompson 1992).
The whole question of the initiation of
MHD jets in association with neutron star formation needs to be
considered anew.

A different mechanism of jet generation involving neutrino radiation
or perhaps MHD jets during collapse of a very
massive star into a black hole has been recently discussed by MacFadyen \&
Woosley (1999) and Woosley (2000) in the context of ``collapsar" models
of \grbs.  The energy input rate and total energy of the jets in
the calculation of MacFadyen \& Woosley
are similar to those of Khokhlov et al. (1999),
but by choice of initial conditions,  MacFadyen \& Woosley inject
energy into the jets as thermal energy whereas Khokhlov et al. assume the
input as kinetic energy.  Apparently this gives less mass in the jets
for MacFadyen \& Woosley and they find that their jets rapidly
accelerate to relativistic speeds, whereas the jets of Khokhlov et al.
remain sub-relativistic.  This affects the dynamics of the jets.
It is as if  MacFadyen \& Woosley were blowing a jet of air
through water and Khokhlov et al. were blowing a jet of water through
water.  If there is to be a strong \grb\ due to Poynting flux
from the neutron star as sketched earlier, there must be
a subsequent phase where, to extend the analogy, a second jet
of air blows out through the water jet.  Clearly, an extensive
amount of work is required to understand the origin of jets in
this general context, their sensitivity to mode of initiation,
and their propagation through various progenitor stars from
compact cores through extended supergiants.

\section{Perspectives and Conclusions}

A principle issue that confronts the subject of \grbs\ is
that of diversity.  Occams razor is a powerful tool, but
sometimes it is not adequate for all of nature's handiwork.
One measure of progress in this century was the development
of our understanding of ``novae."  We now know that this
apparently rather similar category of optical outbursts
included a wide range of astrophysical phenomena: dwarf
novae, classical novae, X-ray transients, and supernovae.
A great amount of painstaking observational work and
theoretical understanding was required to separate these
categories, including the understanding that some of the
``nebulae" hosting ``novae" were galaxies giving rise to
supernovae.  It is sobering to recognize that the differences
between X-ray transients involving black hole accretion
and classical novae involving thermonuclear explosions
on the surfaces of white dwarfs were only fully recognized
in the last decade or so.

One must keep an open mind that
something like this diversity of phenomena may occur in
the \grbs\ despite obvious similarities and in the absence
of other information. That information gap is now being filled
in a rush in the age of the afterglow.  The separation of \grbs\ into
two morphological groups by the length of the outburst
is well established  and into different hardness categories
is suspected (Lamb 1999; Fishman 2000).  The \grbs\ with afterglows
show a variety of light curve behaviors, some with monotonic
power law declines, some with breaks or gradual changes of slope
(Table 2).
Some of the afterglows are seen in the X-ray, but not in the
optical despite simple arguments that say Lorentz beaming
should be more pronounced in higher energy bands.  Time will
tell whether this diversity in phenomenology is telling us
about the complexity of a single category or a diversity
of physical phenomena that share some properties.

SN~1998bw/GRB980425 plays a key role here.  This event (and other
apparently extreme supernovae) lies on the conceptual border
between extreme classes of models.  On one hand, one has
asymmetric supernovae that appear to be commonly associated with
the core collapse phenomenon, Type II and Type Ib/c.  SN~1998bw could
be an extreme version of that physics, requiring energies in
the upper range for other observed supernovae, perhaps
$2\times10^{51}$ ergs versus $\simeq10^{51}$ ergs for most
core collapse events and $1.3\times10^{51}$ ergs for the
well-measured SN~1987A.  In this case, SN~1998bw is predicted
to leave behind a neutron star, a pulsar, perhaps even a
magnetar.  In the other extreme of some order parameter, we
have the classic cosmological \grbs\ as revealed by BeppoSAX and
the sterling follow-up work that has been done in all wavelengths.
These cosmic \grbs\ could be ``hypernovae" involving jets
originating in ``collapsars" as discussed by Woosley (2000) or
neutron star-neutron star or neutron star-black hole collisions
(or the many variations on that theme) or maybe both.  In this
picture, SN~1998bw might represent a mild, not so collimated or
powerful version of the cosmological bursts.  In this case,
SN~1998bw is predicted to leave behind a black hole.  Clearly,
SN~1998bw remains a potential rosetta stone, one of fading,
but still detectable brilliance at the time of the conference.

Too many issues arise in the total sweep of supernova and
\grb\ research to touch on here.  There are some topics that
fall in the interstices that may illuminate both, and
these are worth comment.

One interesting issue is the amplitude of the Lorentz factor.
The Lorentz factor is typically limited to $\lta$ 10 in
both AGN's and the stellar mass black hole sources with
superluminal jets, the microquasars (Mirabel \& Rodriguez 1994;
Fender 1999),
whereas it is understood that that
the Lorentz factors in \grbs\ must be $\gta$ 100 (Baring \& Harding 1997).
What is the difference in the physics of these situations?
One factor that is thought to limit the speed of black hole
jets is radiation drag by photons emitted from the surrounding
disk (Sikora et al. 1996; Luo \& Protheroe 1999).
Does this phenomenon, as well as baryon loading,
affect the models based on black hole accretion? Has
nature, through the AGN and microquasars, already told us
what black hole accretion does, or is the massive accretion
rate postulated for ``collapsar" models sufficient to account for
the difference?  Do \grbs\ with their exceedingly large
Lorentz factors require some physical basis that is very
different than accreting black holes?

In principle, either newly-born neutron stars or black holes
could generate jets.  How are those cases to be differentiated
as we explore SN~1998bw and the other candidate ``hypernova"
events.  One interesting possibility is to look at the
iron abundances in X-ray spectra, as discussed by Piro (2000;
see also M\'esz\'aros \& Rees 1998; Piro et al. 1999a;
Lazzati, Campana \& Ghisellini 1999;
Vietri et al. 1999; Table 2.  Note that a work widely referred to
by Yoshida et al. 1999, based on ASCA observations of GRB~879828,
does not seem to have been submitted).
Another important issue to explore is the nature of the
birth of a ``magnetar" with a magnetic field ranging up to
perhaps $10^{16}$ G compared to the birth of a ``normal"
pulsar with a dipole field of $10^{12}$ G.

The recognition that \grbs\ may involve collimated flows
has become widespread. There is already a debate
about when and at what speed, lateral expansion of jets
will occur and with what effect on the spectrum and
temporal behavior of the afterglow.  The question of
different collimation and Lorentz beaming of the \grb\
and afterglow must be addressed.
Chevalier (2000; Li \& Chevalier 1999; Chevalier \& Li 1999)
has emphasized the
difference of afterglows propagating into density
gradients as opposed to uniform density environments.
In general, a spherical relativistic blast wave propagating
into a density gradient will give a light curve that
declines more steeply than the case of a constant density.
The result is that the effects of density gradients, for instance
winds with density profile $\rho\propto~r^{-2}$, can mimic
the effects of a collimated jet that slows and spreads
in a constant density environment.  In addition, prolonged
energy input into the blast wave can result in a shallower
light curve decline in the case with a density gradient
in a manner that can mimic the effect of a spherical blast wave
in a constant density medium.
The luminosity of a jet could fluctuate in space and
time for dynamical and kinematic reasons.
If one asks about collimated
flow into an environment with density gradients, even clumpiness,
then the range of phenomenology could be quite great
(M\'esz\'aros, Rees \& Wijers 1998).  These issues could be
related.  While relativistic blast waves propagating in a
steady state wind will decelerate, those propagating through
steeper profiles ($\rho \propto r^{-n}$ with n$>$ 3; Shapiro 1980)
will accelerate (Colgate 1975; Shapiro 1980) and the gradient
itself can lead to collimation (Shapiro 1979).
My guess is that the full potential complexity of
the behavior of collimated flow has not yet been
appreciated nor evaluated.

Once again, while it is fading, SN~1998bw could be an important
resource for raising these issues.  All models for this event
presuppose a stripped core of a massive star.  A strong wind
is the most likely candidate for the mass loss and the
radio emission has already given evidence for a relativistic
blast wave interacting with a $\rho\propto~r^{-2}$ density
gradient (Li \& Chevalier 1999).  Follow-up of SN~1998bw in
all possible wavelengths until it is completely unobservable
is strongly encouraged!

Finally, one of the most exciting events in a string of recent revolutions
was the recording of the contemporary optical outburst associated
with GRB~990123 (Akerloff et al. 1999).  This burst was seen at
9th magnitude and Kehoe (2000) emphasized that they cannot
guarantee having detected the peak due to their sampling!
Kehoe also emphasized that they have observed other events at lower
limits so some \grbs, at least, are not this bright.  Still, some obviously
are and the lesson for the ROTSE group and the LOTIS and other
groups who are trying to do automated contemporary detections is
to just be patient.  There must be more such events.

What an incredible event this single observation of GRB~990123
represented.  At
9th magnitude, this optical flare was almost naked eye!  One of the first
things that occurred to me, somewhat facetiously, was encouraging
hoards of private citizens to go out every night in their back
yards with a decent pair of binoculars and look for these events.
BATSE detects about one \grb\ per day.  If every one were like
GRB990123, there should be a bright optical flash for a minute or
so once a day somewhere on the sky that would be easily visible
with a decent pair of binoculars.  A pair of binoculars allows you
to see about one part in a thousand of the total sky.
If you looked every night for three years running, you just
might get lucky.

As it so happens, Gerry Fishman and Janet Mattei of the AAVSO
were way ahead of me.  They are already talking about working
with amateurs with decent sized telescopes and commercial CCD's
to undertake a project akin to Joe Patterson's Backyard Astronomy,
the program he coordinates from Columbia.  With this kind of
organization we may indeed see the era of Backyard Cosmology.

\subsection{Acknowledgments}

My thanks personally and on behalf of all the attendees at the meeting
for a job well done to Mario Livio and the Scientific Organizing
Committee, to Patrick Godon and Rosie Diaz-Miller for handling
the expert and neophyte Power Point operators as well as all
the other A-V tasks, and especially to Lorraine Garcia and
Theresa Bailey for their excellent work on the symposium
arrangements and the myriad issues and questions that always
arise during such a conference.  I am grateful for scientific
discussions of supernovae and gamma-ray bursts with Lifan Wang,
Peter H\"oflich, Rob Duncan, Alexei Khokhlov, Elaine Oran,
Insu Yi, Brian Schmidt, Saul Perlmutter, Alan Sandage, David Branch,
Eddie Baron, Don Lamb, Shri Kulkarni, Josh Bloom,
Dave Meier, Peter M\'eszar\'os,
Martin Rees, Stan Woosley and Andrew MacFadyen.
Special thanks go to Howie Marion for helping with the
data collection and to Martin Lang for LaTeX wrangling.
This research was supported in part by NSF Grant
95-28110, NASA Grant NAG 5-2888, and a grant from the Texas 
Advanced Research Program.


\end{document}